\documentclass[12pt]{article}
\pdfoutput=1
\usepackage{putex}
\usepackage{comment}
\usepackage{graphicx}
\usepackage{caption}
\usepackage{amsmath}
\usepackage{array}
\usepackage{subcaption}
\usepackage{epstopdf}
\usepackage{enumerate}
\usepackage{cite}
\usepackage{youngtab}
\usepackage{tensor}
\usepackage{slashed}
\usepackage[export]{adjustbox}
\usepackage[aligntableaux=center]{ytableau}
\usepackage[utf8]{inputenc}
\usepackage[
      colorlinks=true,
      linkcolor=blue,
      urlcolor=blue,
      filecolor=black,
      citecolor=red,
      ]{hyperref}

\usepackage{braket}

\usepackage{tikz}
\usepackage{subcaption}

\newcommand{\RNum}[1]{\uppercase\expandafter{\romannumeral #1\relax}}

\newcommand {\be} {\begin {equation}}
\newcommand {\ee} {\end {equation}}

\newcommand {\bes} {\begin {equation*}}
\newcommand {\ees} {\end {equation*}}

\newcommand{\beq}{\begin{equation}}
\newcommand{\eeq}{\end{equation}}

\def\eqref#1{(\ref{#1})}

\def\ie{\begin{equation}\begin{aligned}}
\def\fe{\end{aligned}\end{equation}}

\numberwithin{equation}{section}

\def\<{\langle}
\def\>{\rangle}
\def \eps {\epsilon}

\begin{document}


\institution{PU}{Department of Physics, Princeton University, Princeton, NJ 08544, USA}
\institution{UCLA}{Department of Physics \& Astronomy,
University of California, Los Angeles, CA 90095, USA}

\title{On the Large Charge Sector \\ in the Critical $O(N)$ Model at Large $N$}

\authors{Simone Giombi\worksat{\PU}  and Jonah Hyman\worksat{\PU,\UCLA} }

\abstract{We study operators in the rank-$j$ totally symmetric representation of $O(N)$ in the critical $O(N)$ model in arbitrary dimension $d$, in the limit of large $N$ and large charge $j$ with $j/N\equiv \hat{j}$ fixed. The scaling dimensions of the operators in this limit may be obtained by a semiclassical saddle point calculation. Using the standard Hubbard-Stratonovich description of the critical $O(N)$ model at large $N$, we solve the relevant saddle point equation and determine the scaling dimensions as a function of $d$ and $\hat{j}$, finding agreement with all existing results in various limits. In $4<d<6$, we observe that the scaling dimension of the large charge operators becomes complex above a critical value of the ratio $j/N$, signaling an instability of the theory in that range of $d$. Finally, we also derive results for the correlation functions involving two ``heavy" and one or two ``light" operators. In particular, we determine the form 
of the ``heavy-heavy-light" OPE coefficients as a function of the charges and $d$.} 

\maketitle

\tableofcontents

\section{Introduction and Summary}
\label{sec:intro}

Quantum dynamics often simplifies in the limit of large quantum numbers, and results which may be inaccessible within standard perturbation theory can be obtained by a semiclassical calculation.
For example, in the context of the AdS/CFT duality, the expansion at large R-charge \cite{Berenstein:2002jq} and large spin \cite{Gubser:2002tv} has provided many non-trivial tests and crucial insights on the gauge/string duality. Expansions in large quantum numbers have also proved useful in deriving various non-perturbative results in quantum field theory, for example in the context of conformal field theory (CFT), see e.g. \cite{Alday:2007mf, Fitzpatrick:2012yx, Komargodski:2012ek}. Recently, the large charge expansion in CFTs with global symmetry was studied from a rather general viewpoint in \cite{Hellerman:2015nra} using effective field theory methods, see e.g. \cite{Monin:2016jmo, Alvarez-Gaume:2016vff, Jafferis:2017zna, Watanabe:2019pdh, Badel:2019oxl, Antipin:2020abu, Antipin:2020rdw} for further developments, and \cite{Gaume:2020bmp} for a review and a more comprehensive list of references. 

In this note, we study large charge operators in the canonical example of the critical $O(N)$ model in dimension $d$. As it is well-known, this CFT can be described as the IR fixed point (for $d<4$) of the scalar field theory of $N$ fields $\phi^i, i=1,...N$, with the $O(N)$ invariant quartic interaction $\lambda (\phi^i \phi^i)^2$. The IR fixed point can be studied perturbatively in $d=4-\epsilon$  using the Wilson-Fisher $\epsilon$-expansion. Alternatively, it can be studied in general $d$ using the large $N$ expansion. This can be developed by introducing an 
auxiliary field $\sigma$ via the Hubbard-Stratonovich transformation 
\begin{equation}
S = \int d^dx \left(\frac{1}{2}(\partial_{\mu}\phi^i)^2 + \frac{1}{2}\sigma\phi^i \phi^i-\frac{\sigma^2}{4\lambda}\right)\,.
\label{phi4-HS}
\end{equation}
The $1/N$ expansion of the CFT correlation functions can be developed by integrating out the fundamental fields $\phi^i$, which yields an effective action for $\sigma$ where $N$ acts as the coupling constant. In practice, this leads to a set of Feynman diagrammatic rules where one uses an induced $\sigma$ propagator and the $\sigma \phi^i \phi^i$ vertex (see e.g. \cite{Moshe:2003xn, Giombi:2016ejx} for reviews). This standard $1/N$ perturbation theory works as long as one considers correlation functions of operators with quantum numbers that 
are finite in the large $N$ limit. However, when the quantum numbers are of order $N$, the ordinary $1/N$ perturbation theory breaks down. This is because in this case the operator insertions are of the same order as the ``classical" action, and hence the path integral is expected to be dominated by a non-trivial saddle point. In this paper we focus on observables involving scalar operators $O_j$ in the rank-$j$ totally symmetric traceless representation of $O(N)$, in the limit 
\begin{equation}
N\rightarrow \infty\,,\qquad j\rightarrow \infty\,,\qquad \quad {\rm with}~~\frac{j}{N}\equiv \hat{j}\,~~{\rm fixed}\,.
\end{equation}
In this limit, the scaling dimension of the operators are expected to take the form
\begin{equation}
\Delta_j = N h(\hat{j})+O(N^0)
\label{Delj-intro}
\end{equation}
where the non-trivial function $h(\hat{j})$ can be determined by a semiclassical saddle point calculation. In $d=3$, this problem was studied recently in \cite{Alvarez-Gaume:2019biu} using a conformal map to $R_t \times S^2$ (the analogous problem in the $\epsilon$-expansion, where one holds $j\epsilon$ fixed, was studied in \cite{Watanabe:2019pdh, Monin:2016jmo, Antipin:2020abu}). Here we we work in Euclidean $R^d$ throughout, and find the scaling dimensions (\ref{Delj-intro}) for arbitrary $d$. The result is a rather non-trival function of $d$ and $\hat{j}$. It interpolates between a small $\hat{j}$ expansion in integer powers of $\hat{j}$
\begin{equation}
\Delta_j = N \left[\left(\frac{d}{2}-1\right)\hat{j}+h_2(d)\hat{j}^2+\ldots \right]
\label{Del-small}
\end{equation}
and a large $\hat{j}$ expansion of the form
\begin{equation}
\Delta_j = N \hat{j}^{\frac{d}{d-1}}\left[\gamma_0(d)+\frac{\gamma_1(d)}{\hat{j}}+\ldots\right]\,.
\label{Del-large}
\end{equation}
This large $\hat{j}$ behavior is precisely consistent with the effective field theory approach \cite{Hellerman:2015nra, Monin:2016jmo}. Note that, as it is evident from (\ref{Del-small}), this semiclassical evaluation of the scaling dimensions in fact resums an infinite number of terms in the usual $1/N$ expansion, and hence provides an infinite number of checks on standard $1/N$ Feynman diagrams. Having the result for general $d$, we can also make contact with the $\epsilon$-expansion in the overlapping regime of validity with the large $N$ expansion, and we find agreement with all existing results.

To obtain the scaling dimension, we study directly the two-point function of the large charge operators on $R^d$, and determine the semiclassical saddle point for the $\sigma$ field as a function of the insertion points of the ``heavy" operators. This approach also allows us to extract without much further work the correlation functions involving two ``heavy" operators and various ``light" operators. In particular we will derive the expression for the three-point function coefficients in the ``heavy-heavy-light" configuration. Similar results in the effective field theory and analytic bootstrap approaches were previously obtained in \cite{Monin:2016jmo, Jafferis:2017zna}. 

One interesting application of our results is to the $O(N)$ model in $d>4$. It is known that the standard $1/N$ perturbation theory can be formally continued above four dimensions \cite{Parisi:1975im}, and it appears to be unitary and well-defined to all orders in $1/N$ (for operators with quantum numbers that do not scale with $N$). This matches onto the formal UV fixed point of the quartic theory in $d=4+\epsilon$, and onto the IR fixed point of a model with cubic interactions in $d=6-\epsilon$ \cite{Fei:2014yja}. However, as shown in \cite{Giombi:2019upv}, the theory in $4<d<6$ is non-perturbatively unstable due to instanton effects, which lead to small imaginary parts in physical observables. By studying the scaling dimension of large charge operators in this model, here we identify what appears to be another manifestation of the instability of these fixed points. We find that, while the scaling dimensions are real in the small $\hat{j}$ expansion (\ref{Del-small}), the large $\hat{j}$ expansion (\ref{Del-large}) involves complex coefficients. At finite $\hat{j}$, we find that there is a critical value $\hat{j}_{\rm crit}$ such that the scaling dimension is real for $\hat{j}<\hat{j}_{\rm crit}$, and becomes complex for $\hat{j}>\hat{j}_{\rm crit}$. The critical value depends on $d$, and goes to infinity at $d=4$ and $d=6$. In $d=5$, we find the relatively small value $\hat{j}_{\rm crit}\approx 0.052$. Another physical quantity which has been observed to be complex in the critical $O(N)$ model in $4<d<6$ is the thermal free energy on the plane (in other words, the free energy on $S^1\times R^{d-1}$) \cite{Petkou:2018ynm, Giombi:2019upv}. It seems plausible that the latter result is related to the fact that the scaling dimensions of operators with charges of order $N$ are complex. It would be interesting to clarify this further. 

The rest of the paper is organized as follows. In Section \ref{2pt-saddle}, we derive the saddle point equation which determines the semiclassical profile of the Hubbard-Stratonovich field $\sigma$ in the presence of two large charge operators. We then solve the saddle point equation explicitly, and present the final result for the scaling dimension in Section \ref{Delta-sec-final}. We discuss the case of $4<d<6$, where we find complex dimension for sufficiently large charge, in Section \ref{complex-Del}. In Section \ref{corr-func}, we compute correlation functions involving two large charge (``heavy") operators, focusing on the case of 3-point and 4-point functions. Finally, in Section \ref{Concl} we make some concluding remarks and comments on future directions.

\section{The saddle point equation}
\label{2pt-saddle}

Let us consider a scalar composite operator $O_j$ in the spin $j$ totally symmetric traceless 
representation of $O(N)$. A convenient way to describe such operators is to introduce an auxiliary null $N$-component vector $u$, and write $O_j(x,u)=(u\cdot \phi)^j$.\footnote{The tracelessness condition is automatically implemented by the requirement that $u$ is null. One may 
recover the complete traceless symmetric tensor by ``stripping out" the auxiliary 
polarization vectors. This can be done for instance by using a second order differential operator in $u$-space, see e.g. \cite{Dobrev:1975ru}.} For instance, as a special case one may consider the complex combination ${\cal Z}=\phi^1+i\phi^2$, and the operator $O_j={\cal Z}^j$ which carries charge $j$ under the $U(1)\subset O(N)$ (this corresponds to the 
choice $u=(1,i,0,\ldots,0)$).

Conformal symmetry and $O(N)$ symmetry constrains the two-point function of the charge $j$ operators to take the form
\begin{equation}
\langle O_j(u_1,x_1) O_j(u_2,x_2)\rangle =(u_1\cdot u_2)^j \frac{{\cal N}_j}{x_{12}^{2\Delta_j}}
\end{equation}     
where ${\cal N}_j$ is a normalization constant (in general, scheme dependent), and $\Delta_j$ is the scaling dimension that we want to determine. 

The operator $O_j$ is the lowest dimension operator in the sector with charge $j$, and is not expected to undergo mixing. Thus, we can determine the scaling dimension by computing the two-point function as
\begin{equation}
\langle O_j(u_1,x_1) O_j(u_2,x_2)\rangle = \frac{1}{Z} 
\int D\phi D\sigma\, (u_1\cdot \phi(x_1))^j\, (u_2\cdot \phi(x_2))^j
e^{-\int d^dx\left(\frac{1}{2}(\partial_{\mu}\phi^i)^2 + \frac{1}{2}\sigma\phi^i \phi^i\right)}
\end{equation}
where we have introduced the auxiliary Hubbard-Stratonovich field, and dropped the term in (\ref{phi4-HS}) proportional to $\sigma^2/\lambda$ which is irrelevant in the critical limit.\footnote{This is a standard step in developing the $1/N$ perturbation theory of the critical $O(N)$ model. See for instance \cite{Giombi:2016ejx} for a review.} Since the action is quadratic in the $\phi^i$ fields, we may evaluate the two-point function by Wick contractions to get
\begin{equation}
\langle O_j(u_1,x_1) O_j(u_2,x_2)\rangle =j! (u_1\cdot u_2)^j \int D\sigma\, 
\left[G(x_1,x_2;\sigma)\right]^j\, e^{-\frac{N}{2} \log\det\left(-\partial^2+\sigma\right)}\,.
\end{equation}
Here $G(x_1,x_2;\sigma)$ denotes the Green's function of the differential operator $-\partial^2+\sigma$. We are interested in the limit $j\rightarrow \infty$, $N\rightarrow \infty$ with $\hat{j}=j/N$ fixed, and hence we may write
\begin{equation}
\langle O_j(u_1,x_1) O_j(u_2,x_2)\rangle =j! (u_1\cdot u_2)^j \int D\sigma\, 
e^{-N\left(\frac{1}{2} \log\det\left(-\partial^2+\sigma\right)-\hat{j} \log G(x_1,x_2;\sigma)\right)}\,,
\label{OO-sig}
\end{equation}
which highlights the fact that the insertion of the large charge operators contributes a term of order $N$ to the $\sigma$ effective action. In the large $N$ limit, the path integral over $\sigma$ is expected to be dominated by a 
saddle point which extremizes the effective action 
\begin{equation}
S_{\rm eff}[\sigma] = \frac{1}{2} \log\det\left(-\partial^2+\sigma\right)-\hat{j} \log G(x_1,x_2;\sigma)\,.
\label{Seff-sigma}
\end{equation} 
In the absence of the insertion (i.e., $\hat j=0$), the saddle point on $R^d$ is simply $\sigma=0$. However, in the presence of the large charge operators, we expect the saddle point to be at $\sigma(x)=\sigma_*(x;x_1,x_2)$, with $\sigma_*$ a non-trivial profile which depends on the insertion points of the large charge operators. 

To proceed, we make an ansatz for the form of the saddle point profile $\sigma_*$. The key observation is that we may view $\sigma_*$ as the one-point function of $\sigma$ in the presence of the large charge operators. In other words, this is related to the 3-point function $\langle O_j(x_1,u_1) O_j(x_2,u_2)\sigma(x)\rangle$. Following steps similar to the ones above, we have:
\begin{equation}
\frac{\langle O_j(x_1,u_1) O_j(x_2,u_2)\sigma(x)\rangle}{\langle O_j(x_1,u_1) O_j(x_2,u_2)\rangle}=
\frac{\int D\sigma\, \sigma(x)\, e^{-N\left(\frac{1}{2} \log\det\left(-\partial^2+\sigma\right)-\hat{j} \log G(x_1,x_2;\sigma)\right)}}
{\int D\sigma\,  e^{-N\left(\frac{1}{2} \log\det\left(-\partial^2+\sigma\right)-\hat{j} \log G(x_1,x_2;\sigma)\right)}} 
\stackrel{N\rightarrow \infty}{\approx} \sigma_*(x;x_1,x_2)  
\label{sigstar-argument}
\end{equation}
Recalling that $\sigma$ in the critical $O(N)$ model is an operator of scaling dimension $\Delta = 2+O(1/N)$, and using the form of the three-point function of scalar operators fixed by conformal symmetry 
\begin{equation}
    \langle O_1(x_1) O_2(x_2) O_3(x_3)\rangle = \frac{C_{123}}{|x_1-x_2|^{\Delta_1+\Delta_2-\Delta_3} 
|x_1 - x_3|^{\Delta_1+\Delta_3-\Delta_2} |x_2- x_3|^{\Delta_2+\Delta_3-\Delta_1}}
    \label{ONthreepointconformal}
\end{equation}
we deduce that we must have
\begin{equation}
\sigma_*(x;x_1,x_2) = c_{\sigma} \frac{|x_1 - x_2|^2}{|x_1 - x|^2 |x_2- x|^2}\,,
\label{sig-saddle}
\end{equation} 
where $c_{\sigma}$ is an undetermined constant that should be fixed by solving the saddle point equation.\footnote{Note that if we make a conformal transformation to $R_{t}\times S^{d-1}$, and map the insertion points $x_{1,2}$ to $t=\pm \infty$, this maps to a configuration with constant $\sigma_*=c_{\sigma}$ on the 
cylinder $R_t\times S^{d-1}$, as in \cite{Alvarez-Gaume:2019biu}.} Explicitly, this is obtained by extremizing the effective action in (\ref{Seff-sigma}), and reads
\begin{equation}
 \frac{\delta}{\delta \sigma(x)} \left( - \frac{1}{2} \log \det(-\partial^2+\sigma) + \hat{j} \log G(x_1, x_2; \sigma)\right) = 0\,.
 \label{ONsaddlepointequation}
\end{equation}
Computing the functional derivative, one finds
\begin{equation}
    \frac{\delta}{\delta \sigma(x)} \log \det (-\partial^2+\sigma) = G(x, x; \sigma)\,,
    \label{ONfunctionalderivativeequalsgreensfunction}
\end{equation}
and
\begin{equation}
     \frac{\delta}{\delta \sigma(x)} \log(G(x_1, x_2; \sigma)) = - \frac{G(x_1, x; \sigma) G(x, x_2; \sigma)}{G(x_1, x_2; \sigma)}\,.
\label{ONfunctionalderivativeofgreensfunction}
\end{equation}
Combining these two results, we may write the saddle point equation as
\begin{equation}
2 \hat{j}\, G(x_1, x ; \sigma_*) G(x, x_2; \sigma_*) = -G(x,x;\sigma_*)G(x_1, x_2; \sigma_*)\,.
\label{saddle-eq}
\end{equation}
In order to solve for the constant $c_{\sigma}$ in (\ref{sig-saddle}), we will need to evaluate explicitly the Green's function $G(x,y;\sigma_*)$. This is a non-trivial calculation, which we carry out in the next subsection. 

\subsection{The Green's function}
The Green's function is the solution to
\begin{equation}
\left(-\partial^2+\sigma_*(x)\right)G(x,y;\sigma_*) = \delta^d(x-y)
\end{equation}
where $\sigma_*$ is given in (\ref{sig-saddle}). This equation may be solved as a power series in $\sigma_*$, by writing
\begin{equation}
G = G^{(0)} + G^{(1)}+G^{(2)}+\ldots 
\end{equation}
where 
\begin{equation}
\begin{aligned}
&-\partial^2 G^{(0)}(x,y) = \delta^d(x-y)\\
&-\partial^2 G^{(L+1)}(x,y) = -\sigma_*(x) G^{(L)}(x,y)\,,\qquad L=0,1,2,\ldots\,. 
\label{Green-eqs}
\end{aligned}
\end{equation}
Here $G^{(0)}$ is the well-known free field massless propagator 
\begin{equation}
G^{(0)}(x,y) = \frac{C_{\phi}}{|x-y|^{d-2}}\,,\qquad C_{\phi} = \frac{\Gamma\left(\frac{d}{2}-1\right)}{4\pi^{\frac{d}{2}}}\,.
\end{equation}
Solving (\ref{Green-eqs}) iteratively, one then finds
\begin{align}
    \nonumber G^{(L)}&(x, y; \sigma_*) \\
   \nonumber &= (-1)^L C_{\phi}^{L+1}c_{\sigma}^L |x_1 - x_2|^{2L} \int \left( \prod_{k = 1}^L d^d x_k \right) \left( \prod_{k = 1}^L \frac{1}{|x_1 - x_k|^2 |x_2 - x_k|^2} \right) \left( \prod_{j = 0}^{L+1} \frac{1}{|x_{j + 1} - x_{j}|^{d-2}} \right) \\
    &= C_{\phi} g_{12}^L \int \left(\prod_{k=1}^L \frac{d^d x_k}{|x_k - x_1|^2 |(x_k - x_1)-(x_2 - x_1)|^2}  \right) \left( \prod_{j = 0}^{L+1} \frac{1}{|x_{j + 1} - x_{j}|^{d-2}}\textbf{} \right)
\nonumber \\
    &\equiv C_{\phi} D_L \left(x-x_1, y-x_1, x_2 - x_1; 1, 1, \frac{d}{2}-1 \right)
\end{align}
where we defined $g_{12} = -C_{\phi} c_{\sigma}|x_1 -x_2|^2$. Conformal integrals of precisely this kind were evaluated in arbitrary $d$ in \cite{Isaev:2003tk}, exploiting a connection to conformal quantum mechanics. Using the results obtained there, we find
\begin{align}
    \nonumber G(x, y; \sigma_*) &= \sum_{L = 0} ^{\infty} G^{(L)} (x, y; \sigma_*) = C_{\phi} \sum_{L = 0} ^{\infty}D_L \left(x-x_1, y-x_1, x_2 - x_1; 1, 1, \frac{d}{2}-1 \right) \\
\nonumber &= \frac{1}{|x-x_1|^{d-2} |y-x_1|^{d-2}} \sum_{L = 0}^{\infty} \frac{1}{L!} \left( \frac{g_{12}}{4 c_{\phi} |x_1 - x_2|^2} \right)^L \Phi_L (\xi,\eta)\\\
&= \frac{1}{|x-x_1|^{d-2} |y-x_1|^{d-2}} \sum_{L = 0}^{\infty} \frac{1}{L!} \left( -\frac{c_{\sigma}}{4} \right)^L \Phi_L (\xi,\eta)\,
\label{Green-sum}
\end{align}
where\footnote{Our variables $\eta,\xi$ are denoted $u, v$ in \cite{Isaev:2003tk}.}
\begin{equation}
\xi = \frac{x-x_1}{|x-x_1|^2} - \frac{x_1 - x_2}{|x_1 - x_2|^2}\,,\qquad \eta = \frac{y-x_1}{|y-x_1|^2} - \frac{x_1 - x_2}{|x_1 - x_2|^2}
\end{equation}
and 
\begin{equation}
\begin{aligned}
\Phi_L (\xi, \eta) &= -C_{\phi} \int_0 ^{\infty} dt \, t^L \left[ \left( \frac{\xi^2}{\eta^2} \right)^{\alpha} e^{t \alpha}\right]_{\alpha^L} \partial_t \left[ \left( \frac{e^{-t}}{(\xi - e^{-t} \eta)^2} \right)^{\frac{d}{2}-1}\right] \\
&= -\frac{C_{\phi}}{L!} \int_0 ^{\infty} dt \,  \left( t \left( t + \log \frac{\xi^2}{\eta^2} \right) \right)^L \partial_t \left[ \left( \frac{e^{-t}}{(\xi - e^{-t} \eta)^2}\right)^{\frac{d}{2}-1} \right]\,.
\end{aligned}
\end{equation}
Note that $\xi^2 = \frac{|x - x_2|^2}{|x-x_1|^2 |x_2-x_1|^2}$ and $\eta^2 = \frac{|y - x_2|^2}{|y-x_1|^2 |x_2-x_1|^2}$. For $L=0$, we find 
\begin{equation}
\Phi_0 = \frac{C_{\phi}}{|\xi - \eta|^{d-2}} = C_{\phi}\frac{|x-x_1|^{d-2}|y -x_1|^{d-2}}{|x-y|^{d-2}}\,.
\end{equation}
Introducing now the conformal cross ratios
\begin{equation}
    X = \frac{|x-x_1|^2 |y-x_2|^2}{|x_1 - x_2|^2 |x-y|^2}, \,\qquad  Y = \frac{|x-x_2|^2 |y-x_1|^2}{|x_1 - x_2|^2 |x-y|^2}\,,
\label{cross-ratios}
\end{equation}
after an integration by parts, we may write for $L\ge 1$
\begin{equation}
 \Phi_L = \frac{C_{\phi}}{(L-1)!} \frac{|x-x_1|^{d-2} |x-x_2|^{d-2}}{|x-y|^{d-2}} \, \int_0^{\infty} dt \, \frac{\left( t \left( t + \log \frac{Y}{X} \right) \right)^{L-1} \left( 2t + \log \frac{Y}{X} \right)}{\left(1 - X - Y + 2 \sqrt{XY} \cosh\left( t + \frac{1}{2} \log \frac{Y}{X} \right) \right)^{\frac{d}{2}-1}}\,.
\end{equation}
Plugging this into (\ref{Green-sum}) yields (see \cite{Broadhurst:2010ds} for a similar calculation in $d=4$) 
\begin{equation}
\begin{aligned}
 &G(x, y; \sigma_*) \\
    &= \frac{C_{\phi}}{|x-y|^{d-2}} \left(1 + \frac{c_{\sigma}}{4}
    \int_0 ^{\infty} dt \, \frac{(2t + \log \frac{Y}{X})\sum_{L = 1}^{\infty} \frac{(-1)^L}{L! (L-1)!}  \left(\frac{c_{\sigma}}{4} \, t \left( t + \log \frac{Y}{X} \right) \right)^{L-1}}{\left(1 - X - Y + 2 \sqrt{XY} \cosh\left( t + \frac{1}{2} \log \frac{Y}{X} \right) \right)^{\frac{d}{2}-1}} \right) \\
&= \frac{C_{\phi}}{|x-y|^{d-2}} \left(1 
- \frac{\sqrt{c_{\sigma}}}{2}
    \int_0 ^{\infty} dt \, \frac{(2t + \log \frac{Y}{X})J_1 \left( \sqrt{c_{\sigma} t \left(t + \log \frac{Y}{X}\right)}\right)}{\left(1 - X - Y + 2 \sqrt{XY} \cosh\left( t + \frac{1}{2} \log \frac{Y}{X} \right) \right)^{\frac{d}{2}-1}\sqrt{t (t + \log \frac{Y}{X})}}\right) \\
   &= \frac{C_{\phi}}{|x-y|^{d-2}} \left(1 +
    \int_0 ^{\infty} dt\frac{\partial_t \left(J_0 \left(\sqrt{c_{\sigma} t (t + \ell)}\right)\right)}{\left(1 - X - Y+ 2 \sqrt{XY} \cosh \left( t + \frac{\ell}{2} \right) \right)^{\frac{d}{2}-1}} \right)\,,
\label{extra:Gxyintegralgeneral}
\end{aligned}
\end{equation}
where we defined $\ell = \log \frac{Y}{X}$, and $J_k(x)$ denotes the standard Bessel function. After an integration by parts, we finally get 
\begin{equation}
G(x, y; \sigma_*) = \frac{C_{\phi}(d-2)}{|x-y|^{d-2}} \int_0 ^{\infty}dt \frac{ \sqrt{XY} \sinh \left( t + \frac{\ell}{2} \right) J_0 \left(\sqrt{c_{\sigma} t( t + \ell)}\right)}{\left(1 - X - Y + 2 \sqrt{XY} \cosh\left( t + \frac{\ell}{2}\right) \right)^{\frac{d}{2}}}\,.
\label{G-final}
\end{equation}
This is the final result for the Green's function in the presence of the non-trivial profile $\sigma_*$. Note that the fact that it depends on the conformal cross ratios in (\ref{cross-ratios}) is expected from conformal invariance. Indeed, by an argument similar to the one in eq.~(\ref{sigstar-argument}), one can see that the Green's function is related to the four-point function of two ``light" scalar operators in the presence of the two large charge operators. We will come back to this point in section \ref{corr-func} below. 

In order to solve the saddle point equation (\ref{saddle-eq}), we need to evaluate the Green's function (\ref{G-final}) in various limits. Let us first consider the coincident point limit $G(x,x;\sigma_*)$. From (\ref{cross-ratios}), we see that this limit corresponds to $X\rightarrow \infty$, $Y\rightarrow \infty$ and $X/Y\rightarrow 1$, or $\ell\rightarrow 0$. Then we find
\begin{equation}
G(x,x;\sigma_*) = C_{\phi}(d-2)\frac{|x_1-x_2|^{d-2}}{|x-x_1|^{d-2}|x-x_2|^{d-2}}\int_0^{\infty} dt\,\frac{\sinh(t) J_0(\sqrt{c_{\sigma}}t)}{(2\cosh(t)-2)^{\frac{d}{2}}}\,, 
\end{equation}
or, after an integration by parts\footnote{When integrating by parts, we regulate away the power divergence near $t=0$. This is equivalent to evaluating the integral by analytic continuation in $d$.}
\begin{equation}
G(x,x;\sigma_*) = -\frac{C_{\phi}|x_1-x_2|^{d-2}}{|x-x_1|^{d-2}|x-x_2|^{d-2}} \int_0^{\infty} dt\, \frac{\sqrt{c_{\sigma}}J_1(\sqrt{c_{\sigma}}t)}{(2\cosh(t)-2)^{\frac{d}{2}-1}}\,.
\label{Gxx}
\end{equation}
Next, we consider the case when $x \rightarrow x_1$, or $y\rightarrow x_2$, or both. In all of these cases, we have $Y\rightarrow 1$ and $X\rightarrow 0$. Using that
\begin{equation*} 
\sqrt{XY} \sinh \left( t + \frac{\ell}{2} \right) = \frac{1}{2} (Y e^t - X e^{-t}) \text{ and }\sqrt{XY} \cosh \left( t + \frac{\ell}{2} \right) = \frac{1}{2} (Y e^t + X e^{-t}) 
\end{equation*}
the Green's function (\ref{G-final}) may be also written as
\begin{equation}
    G(x, y; \sigma_*) = \frac{C_{\phi}(d-2)}{|x- y|^{d-2}} \int_0^{\infty} dt \,\frac{ \frac{1}{2} (Y e^t - X e^{-t}) J_0 \left( \sqrt{c_{\sigma} t(t + \ell)} \right)} {\left( 1 + X (e^{-t}-1) + Y(e^t-1)\right)^{\frac{d}{2}}}
\end{equation}
Taking the limit $X\rightarrow 0$, $Y\rightarrow 1$ (leaving $\ell$ fixed for now), we have
\begin{equation}
 G(x_1, y; \sigma_*) = \frac{C_{\phi}(d-2)}{2 |x_1- y|^{d-2}} \int_0 ^{\infty} dt \, e^{-t \left( \frac{d}{2} -1 \right)} J_0 \left( \sqrt{c_{\sigma} t(t + \ell)} \right)
\end{equation}
The integral can be evaluated using the identity \cite{gradshteyn2007}
\begin{equation}
    \int_0 ^{\infty} dx \, e^{-\alpha x} J_0 \left( \beta \sqrt{x^2 + 2 \gamma x} \right) = \frac{1}{\sqrt{\alpha^2 + \beta^2}} \exp \left( \gamma \left( \alpha - \sqrt{\alpha^2 + \beta^2} \right) \right)
\end{equation}
which yields the following:
\begin{equation}
    G(x_1, y; \sigma_*) = \frac{C_{\phi}\left( \frac{d}{2}-1 \right)}{|x_1- y|^{d-2} \sqrt{ \left( \frac{d}{2}-1 \right)^2 + c_{\sigma}} }  \; e^{ - \frac{\ell}{2} \left(\sqrt{ \left( \frac{d}{2}-1 \right)^2 + c_{\sigma}}  - \left(\frac{d}{2}-1  \right) \right)}\,.
\label{Glimit-1}
\end{equation}
Now we may plug in the explicit form of $\ell$ in the limit $x\rightarrow x_1$
\begin{equation}
\ell = \log\left(\frac{Y}{X}\right) = \log\left(\frac{|x_1-x_2|^2 |x_1-y|^2}{\delta^2 |x_2-y|^2}\right)\,,
\end{equation}
where we have introduced a small regulator $\delta$ to deal with the short distance singularity that appears when $x$ collides with $x_1$. Plugging this into (\ref{Glimit-1}), we have
\begin{equation}
   G(x_1, y; \sigma_*) = \frac{C_{\phi}(\frac{d}{2}-1)}{|x_1-y|^{d-2}
\sqrt{(\frac{d}{2}-1)^2+c_{\sigma}}}\left(\frac{\delta |x_2-y|}{|x_1-x_2||x_1-y|}\right)^{\sqrt{(\frac{d}{2}-1)^2+c_{\sigma}}-(\frac{d}{2}-1)}\,.
\label{Gx1y}
\end{equation}
Similarly, we have
\begin{equation}
   G(y, x_2; \sigma_*) = \frac{C_{\phi}(\frac{d}{2}-1)}{|x_2-y|^{d-2}
\sqrt{(\frac{d}{2}-1)^2+c_{\sigma}}}\left(\frac{\delta |x_1-y|}{|x_1-x_2||x_2-y|}\right)^{\sqrt{(\frac{d}{2}-1)^2+c_{\sigma}}-(\frac{d}{2}-1)}\,.
\label{Gyx2}
\end{equation}
Finally, when both $x\rightarrow x_1$ and $y\rightarrow x_2$, we get
\begin{equation}
   G(x_1, x_2; \sigma_*) = \frac{C_{\phi}(\frac{d}{2}-1)}{|x_1-x_2|^{d-2}
\sqrt{(\frac{d}{2}-1)^2+c_{\sigma}}}\left(\frac{\delta^2}{|x_1-x_2|^2}\right)^{\sqrt{(\frac{d}{2}-1)^2+c_{\sigma}}-(\frac{d}{2}-1)}\,.
\label{Gx1x2}
\end{equation}

Plugging (\ref{Gxx}), and (\ref{Gx1y})--(\ref{Gx1x2}) into the saddle point equation (\ref{saddle-eq}), we then find the following equation which determines the value of $c_{\sigma}$ at the saddle point
\begin{equation}
\frac{\hat j}{\sqrt{(\frac{d}{2}-1)^2+c_{\sigma}}} =\frac{1}{d-2} \int_0^{\infty}dt\, \frac{\sqrt{c_{\sigma}}J_1(\sqrt{c_{\sigma}}t)}{(2\cosh(t)-2)^{\frac{d}{2}-1}}\,.
\label{saddle-csol}
\end{equation}
This is one of our main results, and it will allow us to obtain the scaling dimensions of the large charge operators by evaluating (\ref{OO-sig}) at the saddle point. The only missing ingredient is the functional determinant, which we obtain in the next subsection. 

\subsection{The functional determinant}
Similarly to the Green's function, the functional determinant may be evaluated as a power series in $\sigma_*$. We have\footnote{The term of order zero in $\sigma$, i.e. $\log\det(-\partial^2)$, is naturally regulated to zero in flat space.}
\begin{equation}
\begin{aligned}
&\log\det(-\partial^2+\sigma_*) \\
&=\sum_{L = 2}^{\infty} \frac{(-1)^{L-1}}{L} \int d^d z_1 \hdots d^d z_L \, \sigma_*(z_1) G_0 (z_1, z_2) \sigma_*(z_2) \hdots G_0(z_{L-1},z_L) \sigma_*(z_L) G_0 (z_L, z_1) \\
&= \sum_{L = 2 }^{\infty} \frac{1}{L} \int d^d z_L \, \sigma_*(z_L)  \left[ (-1)^{L-1} \int d^d z_1 \hdots d^d z_{L-1} \, \sigma_*(z_1) G_0(z_1, z_2) \sigma_*(z_2) \hdots G_0 (z_{L-1}, z_L) G_0 (z_L, z_1)\right] \\
&= \sum_{L=1}^{\infty} \frac{1}{L+1} \int d^dx \, \sigma_*(x) G^{(L)} (x,x; \sigma_*),
\label{logdet}
\end{aligned}
\end{equation}
where for brevity we have omitted the dependence of $\sigma_*$ on the insertion points $x_1, x_2$ of the heavy operators. Expanding (\ref{Gxx}) in powers of $c_{\sigma}$, we can read off
\begin{equation}
 G^{(L)}(x, x; \sigma_*) =C_{\phi}\frac{c_{\sigma}|x_1 - x_2|^{d-2}}{2|x-x_1|^{d-2} |x-x_2|^{d-2}}\int_0^{\infty} dt \, \frac{t}{( 2 \cosh t - 2 )^{\frac{d}{2}-1}} \left[\frac{(-1)^L}{L! (L-1)!} \left( \frac{c_{\sigma}}{4} t^2 \right)^{L-1} \right] 
\end{equation}
Plugging this result into (\ref{logdet}) and using (\ref{sig-saddle}), we find after performing the sum
\begin{equation}
\begin{aligned}
\log\det(-\partial^2+\sigma_*) = 
-2 C_{\phi}  \int d^d x \, \frac{|x_1 - x_2|^{d}}{|x-x_1|^{d} |x-x_2|^{d}} \int_0 ^{\infty} dt \, \frac{c_{\sigma} J_2 \left( \sqrt{c_{\sigma}} t \right)}{t( 2 \cosh t - 2 )^{\frac{d}{2}-1}}\,. 
\end{aligned}
\end{equation}
The integral over $x$ is divergent and needs to be regularized. We will adopt the following analytic regulator 
\begin{equation}
 \int d^d x \, \frac{|x_1 - x_2|^{d}}{|x-x_1|^{d} |x-x_2|^{d}}\rightarrow 
 \int d^{d} x \, \frac{\mu^{2\delta}|x_1 - x_2|^{2(\frac{d}{2}-\delta)}}{|x-x_1|^{2(\frac{d}{2}-\delta)} |x-x_2|^{2(\frac{d}{2}-\delta)}}
\end{equation}
where $\delta\rightarrow 0$ and $\mu$ is a mass scale introduced on dimensional grounds. Using 
\begin{equation}
\int d^dx \frac{1}{|x-x_1|^{2a} |x-x_2|^{2b}} = \pi^{\frac{d}{2}}
\frac{\Gamma\left(\frac{d}{2}-a\right)\Gamma\left(\frac{d}{2}-b\right)\Gamma\left(a+b-\frac{d}{2}\right)}
{\Gamma\left(a\right)\Gamma\left(b\right)\Gamma\left(d-a-b\right)}\frac{1}{|x_1-x_2|^{2(a+b-\frac{d}{2})}}
\end{equation}
we find
\begin{equation}
\int d^{d} x \, \frac{\mu^{2\delta}|x_1 - x_2|^{2(\frac{d}{2}-\delta)}}{|x-x_1|^{2(\frac{d}{2}-\delta)} |x-x_2|^{2(\frac{d}{2}-\delta)}}
=\frac{1}{\delta}\frac{2 \pi^{\frac{d}{2}}}{\Gamma\left( \frac{d}{2} \right)}+\frac{4 \pi^{\frac{d}{2}}}{\Gamma\left( \frac{d}{2} \right)} \log (\mu |x_1 -x_2|)+O(\delta)\,.
\end{equation}
The pole in the regulator that appears here should be removed as part of the renormalization of the composite operator $O_j$. We will drop it in the following and just keep track of the dependence on $\log|x_1-x_2|$, which is sufficient to extract the scaling dimensions. Our final result for the functional determinant is then
\begin{equation}
\log\det(-\partial^2+\sigma_*) = 
 -\frac{4}{d-2}\log (\mu|x_1 - x_2|) \int_0 ^{\infty} dt \, \frac{c_{\sigma}\, J_2 \left( \sqrt{c_{\sigma}} t \right)}{t( 2 \cosh t - 2 )^{\frac{d}{2}-1}}\,.
\label{logdet-final}
\end{equation}

\subsection{The scaling dimension}
\label{Delta-sec-final}
We can now evaluate the two-point function of the large charge operators in the large $N$ limit with $\hat{j}=j/N$ fixed. Using (\ref{OO-sig}), the leading large $N$ result is obtained by evaluating the $\sigma$ effective action (\ref{Seff-sigma}) at the saddle point
\begin{equation}
\langle O_j(x_1,u_1) O_j(x_2,u_2)\rangle \approx  j! (u_1\cdot u_2)^j e^{-N S_{\rm eff}[\sigma_*]}
\label{OO-saddle}
\end{equation}
Let us define
\begin{equation}
f(c_{\sigma}) = - \frac{c_{\sigma}}{d-2} \int_0^{\infty} dt \, \frac{J_2 \left( \sqrt{c_{\sigma}} t \right)}{t(2 \cosh t - 2)^{\frac{d}{2}-1}}\,.
\label{fc}
\end{equation}
Then, using (\ref{Gx1x2}) and (\ref{logdet-final}), we find\footnote{We may identify the short distance cutoff $\delta$ in (\ref{Gx1x2}) to be proportional to $1/\mu$, on dimensional grounds. The proportionality constant is scheme-dependent and does not affect the coefficient of $\log|x_1-x_2|^2$ which is what determines the scaling dimension.}
\begin{equation}
S_{\rm eff}[\sigma_*] = 
\log (\mu^2|x_i - x_f|^2) \left[f(c_{\sigma}) + \hat{j} \sqrt{ \left( \frac{d}{2}-1 \right)^2 + c_{\sigma}}  \right]+ {\rm const.},
\label{Seff-star}
\end{equation}
where $c_{\sigma}$ is the solution of the saddle point equation (\ref{saddle-csol}). Note that, using
\begin{equation}
f'(c_{\sigma}) = -\frac{1}{2(d-2)}\int_0^{\infty}dt \, \frac{\sqrt{c_{\sigma}}J_1(\sqrt{c_{\sigma}}t)}{(2\cosh(t)-2)^{\frac{d}{2}-1}}
\end{equation}
one can see that (\ref{saddle-csol}) is in fact equivalent to simply extremizing the quantity in brackets in (\ref{Seff-star}) with respect to $c_{\sigma}$. 

From (\ref{Seff-star}) and (\ref{OO-saddle}), we can finally read off the scaling dimension to be
\begin{equation}
\begin{aligned}
\Delta_j = N \left[f(c_{\sigma})+\hat{j} \sqrt{ \left( \frac{d}{2}-1 \right)^2 + c_{\sigma}} \right]_{c_{\sigma}=c_{\sigma}(\hat{j})} 
\end{aligned}
\label{Delta-final}
\end{equation}
where $c_{\sigma}(\hat{j})$ is the solution to (\ref{saddle-csol}), or equivalently
\begin{equation}
    \frac{d}{dc_{\sigma}} \left[f(c_{\sigma}) + \hat{j} \sqrt{ \left( \frac{d}{2}-1 \right)^2 + c_{\sigma}}  \right] = 0\,.
    \label{saddle-csol-extra}
\end{equation}

\begin{figure}[ht]
\begin{center}
\includegraphics[width=4in]{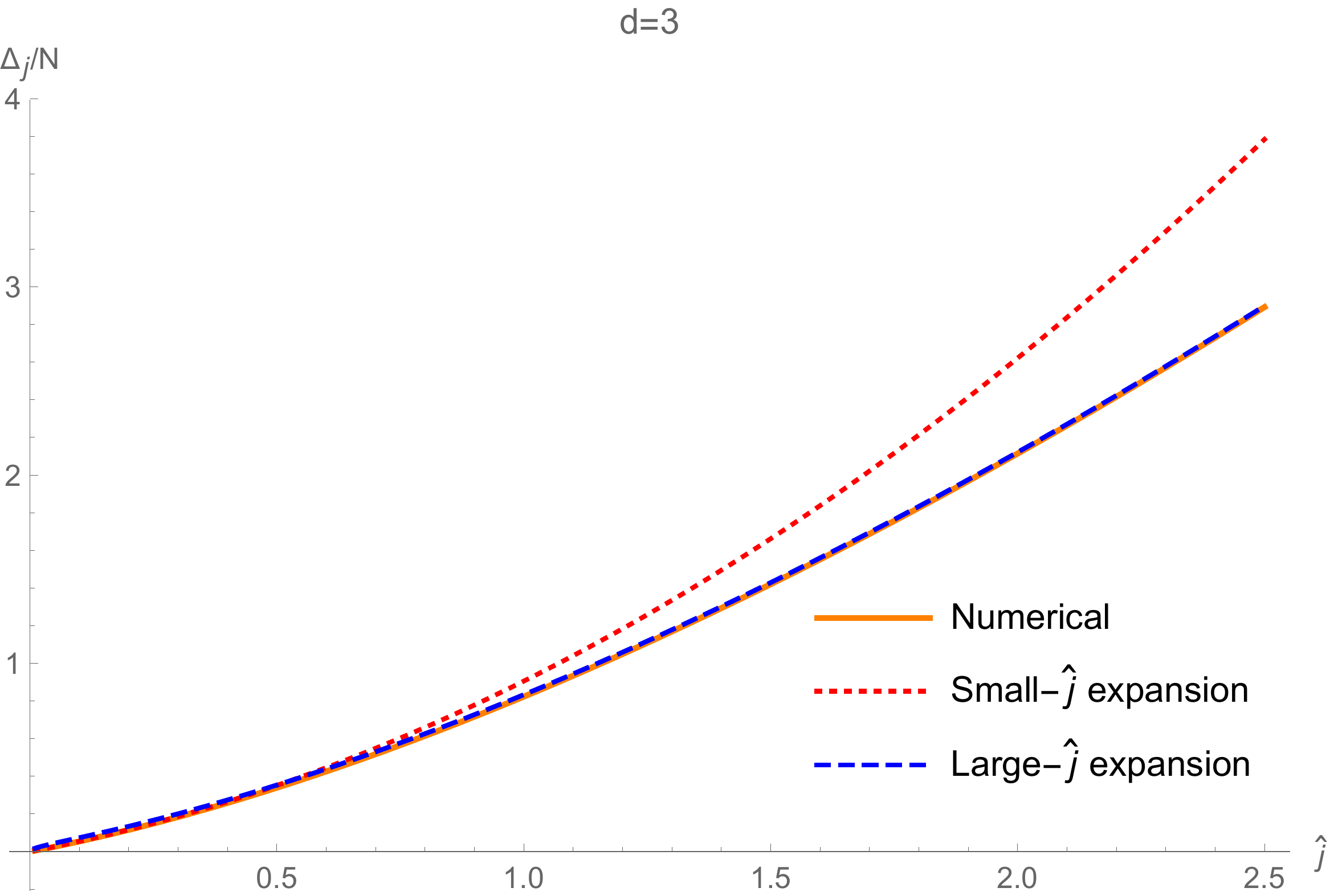}
\caption{The scaling dimension $\Delta_j/N$ as a function of $\hat{j}=j/N$ in $d=3$. The solid line was obtained by numerically solving (\ref{saddle-csol-extra}). The small and large $\hat{j}$ curves correspond to eq.~(\ref{3d-small-j}) and (\ref{3d-large-j}), keeping the first two terms in each expansion. The analytic large $\hat{j}$ expansion is remarkably close to the exact answer even for relatively small $\hat{j}$.}
\label{Fig:delj-3d}
\end{center}
\end{figure}

This equation may be solved numerically for finite $\hat{j}$, or analytically in the small $\hat{j}$ or large $\hat{j}$ expansions, as we describe below. In figure \ref{Fig:delj-3d} we plot the scaling dimension as a function of $\hat{j}$ in $d=3$, obtained by numerically solving (\ref{saddle-csol-extra}), and compare it to the analytic expansions at small and large $\hat{j}$. 

\paragraph{Small $\hat{j}$ expansion} 
In the small $\hat{j}$ limit, we may solve (\ref{saddle-csol-extra}) in powers of $\hat{j}$. Note that $c_{\sigma}\rightarrow 0$ as $\hat{j}\rightarrow 0$. Using 
\begin{equation}
f(c_{\sigma}) = \sum_{k=0}^{\infty} \frac{\left( - \frac{1}{4} \right)^{k+1}}{(d-2) k! (k+2)!} c_{\sigma}^{k+2} I_k(d)\,,
\end{equation}
where 
\begin{equation}
  I_k(d) = \int _0^{\infty} dt\,\frac{ t^{2k +1}}{(2 \cosh t - 2)^{\frac{d}{2}-1}} =
(2k+1)! \sum_{j=0}^{\infty} \frac{(-1)^j  \binom{2-d}{j}}{\left(\frac{d}{2}+j-1\right)^{2k+2}} \,,
\end{equation}
we find 
\begin{equation}
\Delta_j = N \left[\left(\frac{d}{2}-1\right)\hat{j}+h_2(d)\hat{j}^2+h_3(d)\hat{j}^3+\ldots \right]
\label{Delta-smallj}
\end{equation}
where
\begin{equation}
\begin{aligned}
&h_2(d) =-\frac{2^{d-3} d \sin \left(\frac{\pi  d}{2}\right) \Gamma \left(\frac{d-1}{2}\right)}{\pi ^{3/2} \Gamma \left(\frac{d}{2}+1\right)}\\
&h_3(d)= -\frac{(d-2) d^2 \Gamma (d-2)^2 \left(\pi ^2-6 \psi ^{(1)}\left(\frac{d}{2}\right)\right)}{6 \Gamma \left(2-\frac{d}{2}\right)^2 \Gamma \left(\frac{d}{2}-1\right)^4 \Gamma \left(\frac{d}{2}+1\right)^2}
\label{h2h3}
\end{aligned}
\end{equation}
and the higher order coefficients are straightforward to obtain, though they become rather lengthy. Note that, recalling that $\hat{j}=j/N$, the expression in (\ref{Delta-smallj}) contains an infinite number of terms from the point of view of the usual $1/N$ expansion, namely those with the highest power of $j$ at each order in $1/N$. The expression for $h_2(d)$ can be seen to be in agreement with the known result for the anomalous dimension of the charge-$j$ operators to order $1/N$ \cite{Lang:1992zw}, which can be computed by standard Feynman diagram methods 
\begin{equation}
\Delta_j = \left(\frac{d}{2}-1\right)j -\frac{2^{d-3} d \sin \left(\frac{\pi  d}{2}\right) \Gamma \left(\frac{d-1}{2}\right)}{\pi ^{3/2} \Gamma \left(\frac{d}{2}+1\right)}\frac{j(j-2)+\frac{4j}{d}}{N}+O\left(\frac{1}{N^2}\right)\,.
\end{equation}
The correction of order $1/N^2$ was also computed in \cite{Derkachov:1997ch}, and one can check that the term of order $j^3/N^2=N \hat{j}^3$ in the result obtained there precisely matches the function $h_3(d)$ in (\ref{h2h3}).\footnote{There is a typo in eq.~(5.23) of \cite{Derkachov:1997ch}: the first term in the bracket should be multiplied by a factor of $(\mu-1)^2=(d/2-1)^2$. We thank A. Manashov for pointing this out to us.}

Specializing (\ref{Delta-smallj}) to $d=3$, one finds
\begin{equation}
\Delta_j =  N\left[\frac{\hat{j}}{2}+\frac{4}{\pi ^2}\hat{j}^2+\frac{16 \left(\pi ^2-12\right)}{3 \pi ^4}\hat{j}^3+
\frac{16 \left(384-48 \pi ^2+\pi ^4\right)}{3 \pi ^6}\hat{j}^4+\ldots\right]\,,
\label{3d-small-j}
\end{equation}
which agrees with the result found in \cite{Alvarez-Gaume:2019biu}. Similarly, in $d=5$ we get
\begin{equation}
\Delta_j = N\left[\frac{3}{2}\hat{j}  -\frac{32}{3 \pi ^2}\hat{j}^2+\frac{1024 \left(3 \pi ^2-40\right)}{27 \pi ^4}\hat{j}^3
-\frac{8192 \left(13520-1548 \pi ^2+27 \pi ^4\right)}{243 \pi ^6}\hat{j}^4+\ldots\right]\,.
\label{5d-small-j}
\end{equation}
It is also useful to compare our result to the $\epsilon$-expansion. Setting $d=4-\epsilon$, we find, working up to order $\epsilon^3$
\begin{equation}
\Delta_j =N\left[\left(1-\frac{\epsilon}{2}\right)\hat{j}
+ \left(\epsilon-\frac{\epsilon^2}{2}-\frac{\epsilon^3}{4}+\ldots\right)\hat{j}^2
+\left(-2\epsilon^2+\epsilon^3(2\zeta_3+1)+\ldots \right)\hat{j}^3
+\left(8\epsilon^3+\ldots \right)\hat{j}^4+\ldots\right]\,.
\end{equation}
This is in precise agreement with the result obtained long ago in \cite{Wallace_1975}.\footnote{In \cite{Wallace_1975} the result is written in 
terms of the exponent $\alpha(j)$, which is related to $\Delta_j$ by $2\alpha(j)=\Delta_j - j \Delta_1$, where $\Delta_1$ is the scaling dimension of the fundamental field.} 

\paragraph{Large $\hat{j}$ expansion} 
To obtain the expansion of the scaling dimension at large $\hat{j}$, one may rescale the integration variable in (\ref{fc}) and expand in inverse powers of $c_{\sigma}$. Using the integral 
\begin{equation}
    \int_0 ^{\infty} dx \, x^k J_2(x) = 2^k \frac{\Gamma \left( \frac{3 + k}{2} \right)}{\Gamma \left( \frac{3-k}{2} \right)}
\end{equation}
and solving (\ref{saddle-csol-extra}) order by order in $1/\hat{j}$, we get
\begin{equation}
    c_{\sigma} = \hat{j}^{\frac{2}{d-1}} C_0(d) + C_1(d) + \frac{C_2(d)}{\hat{j}^{\frac{2}{d-1}}} + \ldots
\end{equation}
where
\begin{equation}
\begin{aligned}
&C_0 = \left( -\frac{2^d}{\pi d}\sin\left( \frac{\pi d}{2} \right) \Gamma\left( \frac{d}{2} \right) \Gamma \left( 1 + \frac{d}{2} \right) \right)^{\frac{2}{d-1}}\\
&C_1 = - \frac{1}{12} (d-2)^2\\
&C_2 =  \frac{1}{360} (d-2)^2 (d^2-3d+6) \frac{1}{C_0}\,.
\label{C0}
\end{aligned}
\end{equation}
Plugging this expansion back into equation (\ref{Delta-final}), we get
\begin{equation}
    \Delta_j = N \hat{j}^{\frac{d}{d-1}}\left(\gamma_0 +  \frac{\gamma_1}{\hat{j}}+  \frac{\gamma_2}{\hat{j}^2}+ \ldots \right) 
\label{del-largej}
\end{equation}
where
\begin{equation}
\begin{aligned}
&\gamma_0 = \left(1-\frac{1}{d}\right)(C_0)^{\frac{1}{2}}\,,\qquad 
\gamma_1 = \frac{(d-1)(d-2)}{12} (C_0)^{-\frac{1}{2}}\\
&\gamma_2 = -\frac{(d-1)(d-2)^2 (3d-2)}{1440} (C_0)^{-\frac{3}{2}}\,.
\end{aligned}
\end{equation}
Note that the large $j$ behavior $\Delta_j \sim j^{\frac{d}{d-1}}$ agrees with the prediction of the effective field theory approach \cite{Hellerman:2015nra, Monin:2016jmo}. 

For $d = 3$, eq.~(\ref{del-largej}) reduces to 
\begin{equation}
    \Delta_j = N \left(\frac{2}{3} \hat{j}^{\frac{3}{2}} + \frac{1}{6} \hat{j}^{\frac{1}{2}} -\frac{7}{720}\frac{1}{\hat{j}^{\frac{1}{2}}} -\frac{71}{181440}\frac{1}{\hat{j}^{\frac{3}{2}}}+\ldots\right)
\label{3d-large-j}
\end{equation}
which agrees with \cite{Alvarez-Gaume:2019biu}. In $d=4-\epsilon$, we get
\begin{equation}
\Delta_j =  \frac{3N}{2^{\frac{4}{3}}} \epsilon^{\frac{1}{3}} \hat{j}^{\frac{4}{3}}+\ldots\,.
\label{eps-exp}
\end{equation}
This can be seen to be in agreement with the result obtained in \cite{Antipin:2020abu}, taking the appropriate large $N$ limit. 

\subsection{Complex dimensions in $4<d<6$}
\label{complex-Del}
Note that if we try to continue eq.~(\ref{eps-exp}) to $d=4+\epsilon$, due to the fractional power of $\epsilon$ the scaling dimension in the large $\hat{j}$ limit becomes complex
\begin{equation}
\Delta_j \approx e^{\pm i \frac{\pi}{3}}\frac{3N}{2^{4/3}}   \eps^{1/3}\hat{j}^{4/3}\,,\qquad d=4+\epsilon\,.
\label{Deltaj-4pluseps}
\end{equation} 
In fact, one can see that the large $\hat{j}$ expansion (\ref{del-largej}) yields complex dimensions in the whole range $4<d<6$, and in particular in $d=5$, because $C_0$ in (\ref{C0}) is complex in that range of $d$. Explicitly, in $d=5$ eq.~(\ref{del-largej}) yields
\begin{equation}
\Delta_j = e^{\pm i\frac{\pi}{4}} \frac{4\sqrt{3}N}{5}\hat{j}^{\frac{5}{4}}+\ldots\,,\qquad d=5\,. 
\end{equation}
Similary, the scaling dimension at large $\hat{j}$ is complex in $d=6-\eps$, where one finds $\Delta_j \approx -e^{\pm i4\pi/5}\frac{5N}{3}(2\eps)^{1/5}\hat{j}^{6/5}$.\footnote{The scaling dimensions of large charge operators in the IR fixed point of the cubic model in $d=6-\epsilon$ were recently studied in \cite{Arias-Tamargo:2020fow}. However, that work focuses on the regime $j\sqrt{\epsilon}$ fixed, which corresponds to small $j\epsilon$. To detect the complex dimensions, one would need to study the limit of large $j\epsilon$.} 
On the other hand, the small $\hat{j}$ expansion (\ref{Delta-smallj}) still yields real scaling dimensions in $4<d<6$ (see also (\ref{5d-small-j}) for the case of $d=5$). This suggests that there is a critical value of $\hat{j}$ at which a real solution to the saddle point equation ceases to exist, and the scaling dimensions become formally complex for $\hat{j}>\hat{j}_{\rm crit}$. Physically, the appearance of complex dimensions should be interpreted as a manifestation of an instability of the theory, which is detected in the sector of operators with charge of order $N$. Note that in $d=4+\epsilon$, the fact that the dimension in (\ref{Deltaj-4pluseps}) is complex can be seen to be directly related to the fact that the fixed point coupling in the $\phi^4$ theory is negative.\footnote{The result of \cite{Badel:2019oxl, Antipin:2020abu}, in the limit of large $j\lambda_*$, takes the form $\Delta_j \approx \frac{3(j\lambda_*)^{4/3}}{2(4\pi)^{2/3}\lambda_*} $, where $\lambda_*$ is the fixed point coupling of the $\lambda/4(\phi^i\phi^i)^2$ theory. In $d=4-\epsilon$, it is given by $\lambda_*=8\pi^2\epsilon/(N+8)+O(\epsilon^2)$.}

\begin{figure}
\begin{center}
\includegraphics[width=4in]{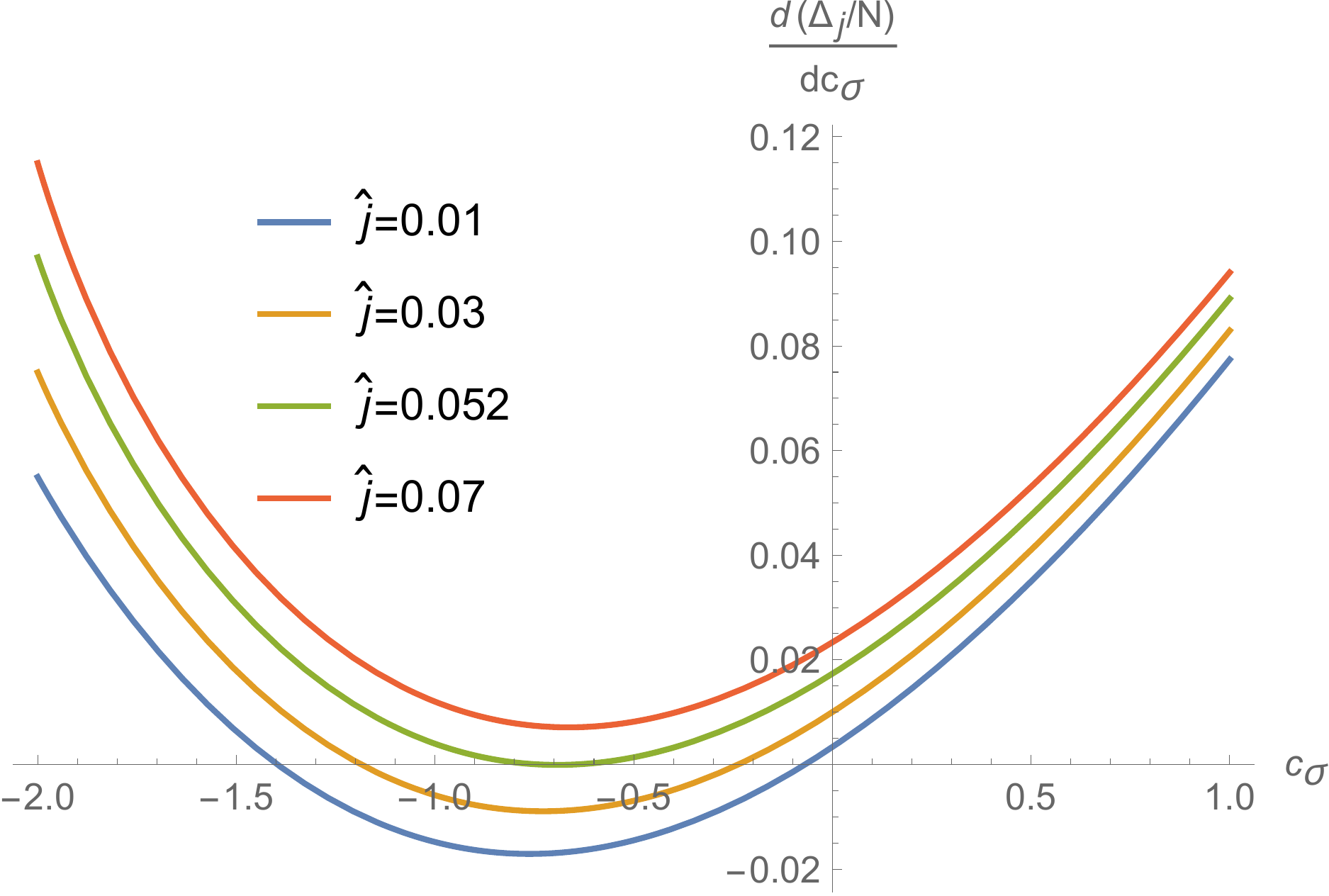}
\caption{Plot of the left-hand side of the saddle point equation (\ref{saddle-csol-extra}) as a function of $c_{\sigma}$ in $d=5$, for several values of $\hat{j}=j/N$. There are no real solutions for $\hat{j}>\hat{j}_{\rm crit}\simeq 0.052$.} 
\label{Fig:csig-5d-saddle}
\end{center}
\end{figure}

\begin{figure}
\begin{center}
\includegraphics[width=4in]{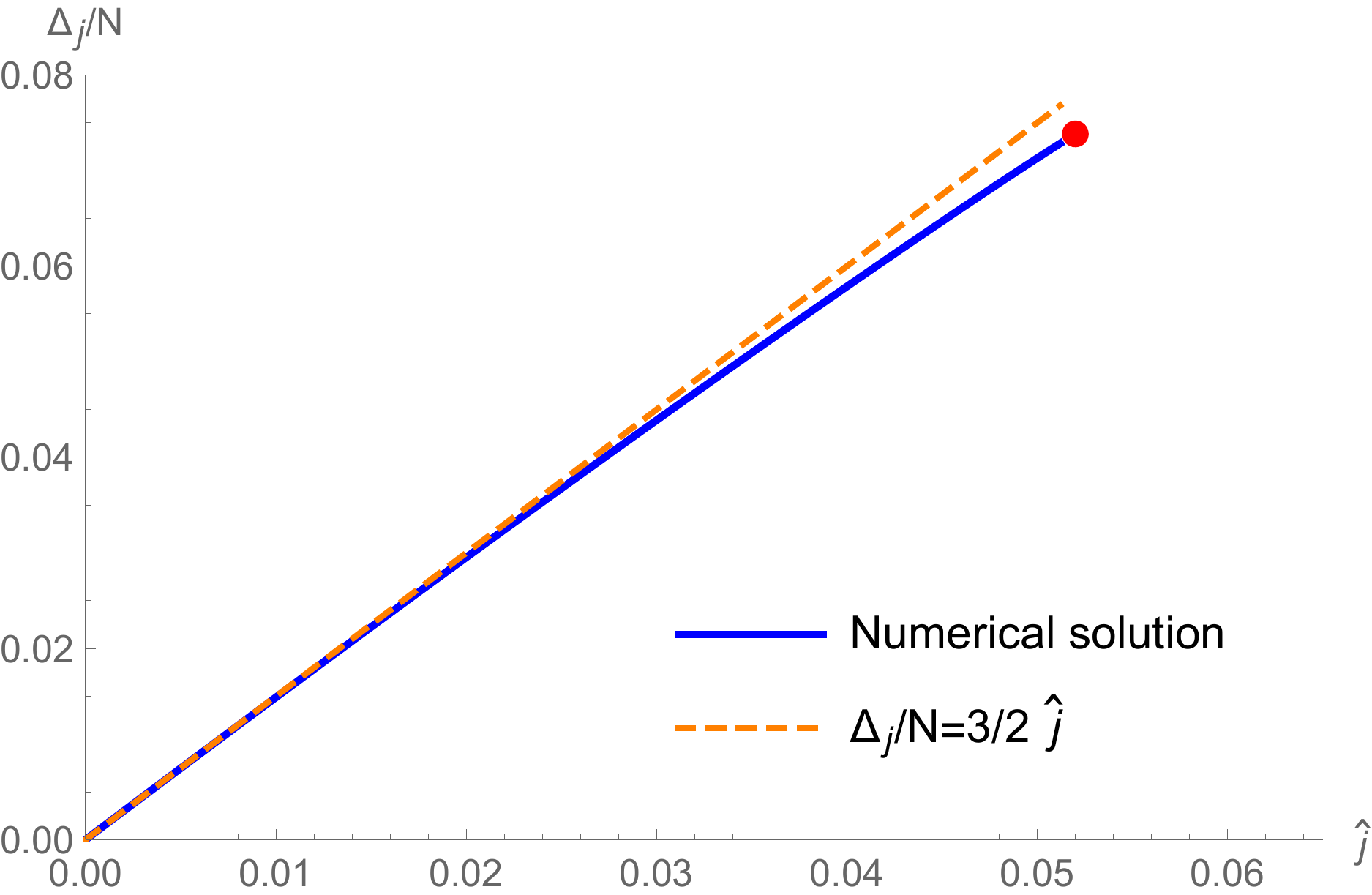}
\caption{Scaling dimension in $d=5$ as a function of $\hat{j}=j/N$. The solid line corresponds to the numerical solution of the saddle point equation, and the dashed line to the small $\hat{j}$ approximation $\Delta_j = \frac{3N}{2}\hat{j}$. the red dot corresponds to the critical values $\hat{j}_{\rm crit} \simeq 0.052$, $\Delta_{j_{\rm crit}}/N \simeq 0.074$. The scaling dimension becomes complex for $\hat{j}>\hat{j}_{\rm crit}\simeq 0.052$.} 
\label{Fig:delj-5d-saddle}
\end{center}
\end{figure}

\begin{figure}
\begin{center}
\includegraphics[width=4in]{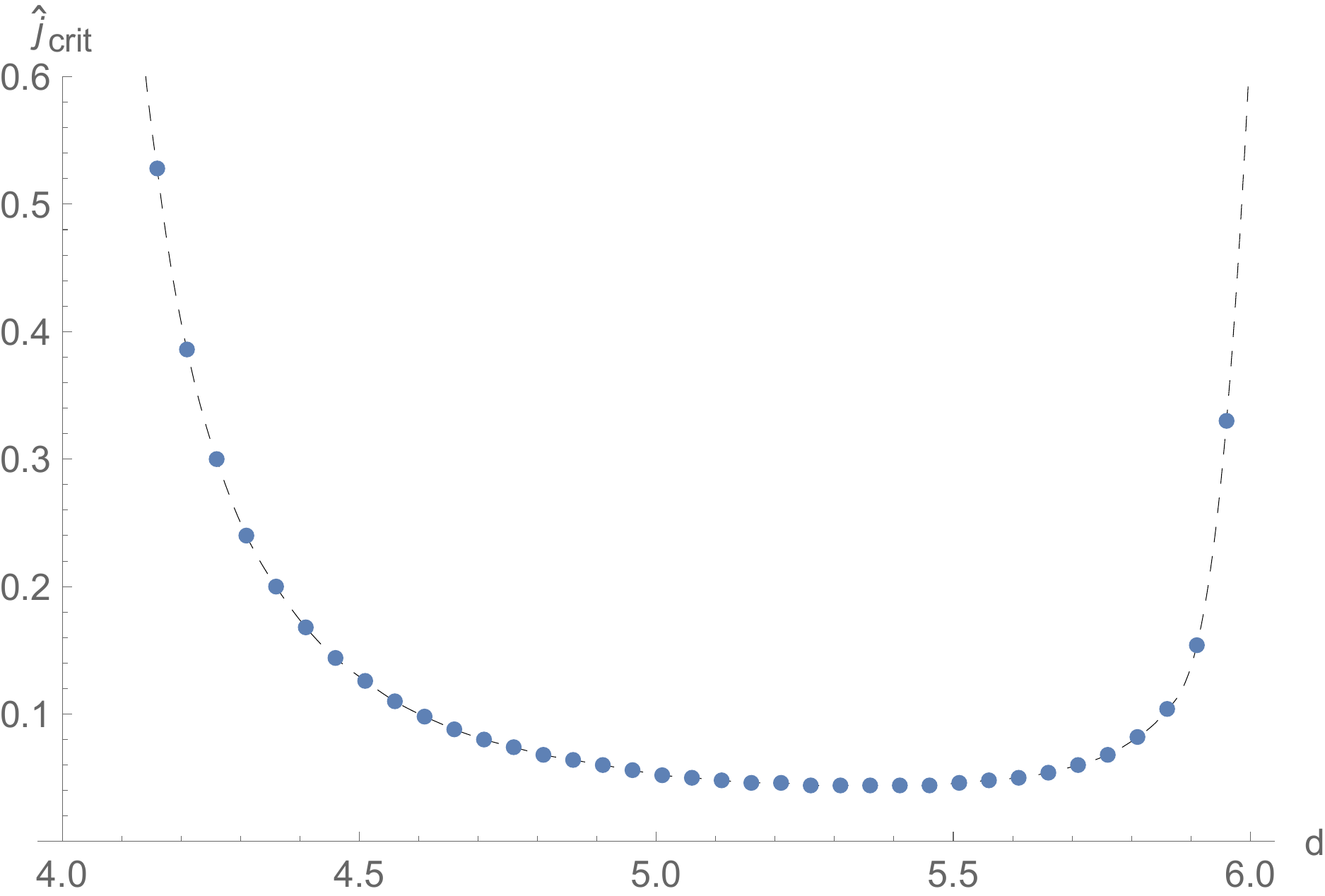}
\caption{Numerical estimate of the critical value of $\hat{j}$ as a function of $d$, for $4<d<6$. The dashed line is a smooth interpolation of the numerical results.} 
\label{Fig:jcritvsd}
\end{center}
\end{figure}

We can estimate the value of $\hat{j}_{\rm crit}$ by numerically analyzing the saddle point equation. In figure \ref{Fig:csig-5d-saddle} we plot the left-hand side of eq.~(\ref{saddle-csol-extra}) in $d=5$, for several values of $\hat{j}$. We see that for small $\hat{j}$ there are two real solutions to the saddle point equation (the solution with largest $c_{\sigma}$ is the one that is smoothly connected to the expected small $\hat{j}$ expansion where $\Delta_j=3N/2\hat{j}+\ldots$). As $\hat{j}$ is increased, the two solutions get closer and ``collide" at 
$\hat{j}\simeq 0.052$, and then move off to the complex plane for $\hat{j}> 0.052$. This mechanism is qualitatively rather similar to what happens in the so-called complex CFTs \cite{Gorbenko:2018ncu}. In figure \ref{Fig:delj-5d-saddle} we plot the scaling dimension as a function of $\hat{j}$, up to the point where the real solution ceases to exist, which corresponds to
\begin{equation}
\hat{j}_{\rm crit} \simeq 0.052\,,\qquad \Delta_{j_{\rm crit}} \simeq 0.074N\,.
\end{equation}

In a similar way, we can estimate the value of $\hat{j}_{\rm crit}$ as a function of $d$, in the range $4<d<6$. This is shown in figure \ref{Fig:jcritvsd}. The smooth interpolation of the numerical results is consistent with $\hat{j}_{\rm crit}\rightarrow \infty$ at $d=4$ and $d=6$, where the critical $O(N)$ model becomes a free CFT. Interestingly, the function $\hat{j}_{\rm crit}(d)$ appears to be qualitatively similar to the function $f(d)$ controlling the instanton induced imaginary parts $\sim e^{-N f(d)}$ that was found in \cite{Giombi:2019upv}. It would be interesting to clarify the relation between these quantities.

\section{Correlation functions at large charge}
\label{corr-func}
Having obtained the Green's function (\ref{G-final}) as a function of the insertion points of the large charge operators, it is relatively straightforward to derive the correlation functions of two ``heavy" and an arbitrary number of 
``light" operators. Below we focus on three-point functions, from which we can extract the OPE coefficients in the ``heavy-heavy-light" configuration, and on the four point functions in the ``heavy-heavy-light-light" configuration. 

\subsection{Three-point functions}
The three-point function of scalar operators with charges $j_1$, $j_2$, $j_3$ (in the totally symmetric traceless representation of $O(N)$) is fixed by conformal symmetry and $O(N)$ symmetry to take the form
\begin{equation}
\langle O_{j_1}(x_1,u_1)O_{j_2}(x_2,u_2)O_{j_3}(x_3,u_3) \rangle=
C_{j_1 j_2 j_3} \frac{(u_1\cdot u_2)^{(j_1+j_2-j_3)/2} 
(u_1 \cdot u_3)^{(j_1+j_3-j_2)/2} (u_2\cdot u_3)^{(j_2+j_3-j_1)/2}}
{|x_{12}|^{\Delta_{j_1}+\Delta_{j_2}-\Delta_{j_3}} 
|x_{13}|^{\Delta_{j_1}+\Delta_{j_3}-\Delta_{j_2}} 
|x_{23}|^{\Delta_{j_2}+\Delta_{j_3}-\Delta_{j_1}}}\,.
\end{equation}
The $O(N)$ symmetry requires this 3-point function to vanish unless the charges satisfy the triangular inequalities $j_i+j_j\ge j_k$, and $\sum_i j_i = {\rm even}$. 

Let us now consider the heavy-heavy-light configuration
\begin{equation}
\begin{aligned}
&j_1 = j+q\,,\quad j_2=j\,, \quad j_3, q{\rm ~~fixed}\\
&j\rightarrow \infty\,,\quad {\rm with} ~~ \hat{j}=j/N{\rm ~~fixed}\,.
\end{aligned}
\end{equation}
Note that $O(N)$ symmetry requires $-j_3\leq q \leq j_3$ (and $j_3+q={\rm even}$). Now using the explicit form of the operators $O_j(x,u)=(u\cdot \phi(x))^j$, we have
\begin{equation}
\begin{aligned}
&\langle O_{j_1}(x_1,u_1)O_{j_2}(x_2,u_2)O_{j_3}(x_3,u_3) \rangle \\
&= \frac{1}{Z}\int D\phi D\sigma\, (u_1\cdot \phi(x_1))^{j+q}\, (u_2\cdot \phi(x_2))^j (u_3\cdot \phi(x_3))^{j_3} 
e^{-\int d^dx\left(\frac{1}{2}(\partial_{\mu}\phi^i)^2 + \frac{1}{2}\sigma\phi^i \phi^i\right)}\\
& =n_{j+q,j,j_3} \int D\sigma\, 
\left(u_1\cdot u_2 G_{12}\right)^{j+(q-j_3)/2} 
\left(u_1\cdot u_3 G_{13}\right)^{(j_3+q)/2} 
\left(u_2\cdot u_3 G_{23}\right)^{(j_3-q)/2} 
\, e^{-\frac{N}{2} \log\det\left(-\partial^2+\sigma\right)}\,,
\label{HHL-3pt}
\end{aligned}
\end{equation}
where we used the shorthand $G_{ij}=G(x_i,x_j;\sigma)$, and the $n_{j+q,j,j_3}$ factor comes from the combinatorics of Wick contractions, which gives
\begin{equation}
n_{j_1,j_2,j_3} = \frac{j_1! j_2! j_3!}
{(\frac{j_1+j_2-j_3}{2})!
(\frac{j_1+j_3-j_2}{2})!
(\frac{j_2+j_3-j_1}{2})!}\,.
\end{equation}
Now we note that in (\ref{HHL-3pt}), the only term that affects the calculation of the saddle point at large $N$ is the factor $G_{12}^j = \exp(N \hat{j}\log(G_{12}))$, which is the same as in the two-point function calculation in section \ref{2pt-saddle}. Therefore, the $\sigma$ path-integral is dominated by the same saddle point as found there, and we simply have to evaluate all factors in (\ref{HHL-3pt}) at $\sigma=\sigma_*$. Stripping off the position dependent and polarization dependent factors which are fixed by symmetry, this yields for the 3-point function coefficient
\begin{equation}
C_{j+q,j,j_3} =n_{j+q,j,j_3}\,{\cal N}^j {\cal N}^{\frac{j_3+q}{2}}\,,\qquad  \quad {\cal N} \equiv  \frac{C_{\phi}(\frac{d}{2}-1)}{\sqrt{(\frac{d}{2}-1)^2+c_{\sigma}}}\,,
\label{C3}
\end{equation}
where ${\cal N}$ is the normalization factor coming from the Green's function, see eqs.~(\ref{Gx1y})--(\ref{Gx1x2}). To obtain the 3-point coefficient for unit normalized operators, which we denote by $a_{j+q,j,j_3}$, we may divide (\ref{C3}) by the square root of the two-point function normalization factors. This yields
\begin{equation}
a_{j+q,j,j_3} = \frac{C_{j+q,j,j_3}}{\sqrt{(j+q)! {\cal N}^{j+q} j! {\cal N}^j j_3! C_{\phi}^{j_3}}} 
= \frac{n_{j+q,j,j_3}}{\sqrt{(j+q)! j! j_3!}} \left(\frac{\frac{d}{2}-1}{\sqrt{(\frac{d}{2}-1)^2+c_{\sigma}}}\right)^{\frac{j_3}{2}}\,.
\end{equation}
Recalling that we are working in the large $N$ limit with $j/N=\hat{j}$ fixed, we obtain the final result
\begin{equation}
a_{j+q,j,j_3} =  \frac{\sqrt{j_3!}}{\left(\frac{j_3+q}{2}\right)!\left(\frac{j_3-q}{2}\right)!}\,
\left(\frac{(\frac{d}{2}-1)N\hat{j}}{\sqrt{(\frac{d}{2}-1)^2+c_{\sigma}}}\right)^{j_3/2}\,,
\label{a-final}
\end{equation}
where $c_{\sigma}$ is fixed in terms of $\hat{j}$ by (\ref{saddle-csol-extra}). In particular, in the limit of large $\hat{j}$, we get 
\begin{equation}
a_{j+q,j,j_3} = N^{j_3/2}\, \frac{\sqrt{j_3!}}{\left(\frac{j_3+q}{2}\right)!\left(\frac{j_3-q}{2}\right)!}\,
\left(\frac{d/2-1}{\sqrt{C_0}}\right)^{j_3/2} \hat{j}^{\frac{(d-2)j_3}{2(d-1)}}\left[1-\frac{1}{24}(d-2)^2 \frac{j_3}{C_0 \hat{j}^{2/(d-1)}}+\ldots \right]\,,
\end{equation}
with $C_0$ given in (\ref{C0}). The leading large $\hat{j}$ scaling $a_{j+q,j,j_3}\sim j^{\frac{(d-2)j_3}{2(d-1)}} = j^{\Delta_{j_3}/(d-1)}$ agrees in $d=3$ with the EFT result obtained in \cite{Monin:2016jmo} (see also \cite{Jafferis:2017zna}).  

\subsection{Four-point functions}
For simplicity, let us specialize to the case of four-point functions of two large charge operators and two fundamental (charge 1) fields. Also, let us split $\phi^i = (\phi^1,\phi^2,\varphi^a), a=1,\ldots, N-2$, and take the ``heavy" operators to be ${\cal Z}^j = (\phi_1+i \phi_2)^j$ and $\bar{\cal Z}^j=(\phi_1-i\phi_2)^j$. Then we can consider two kinds of heavy-heavy-light-light 4-point functions: $\langle {\cal Z}^j\, \bar{\cal Z}^j\,\varphi^a\, \varphi^b\rangle$ and $\langle {\cal Z}^j\, \bar{\cal Z}^j\, {\cal Z}\, \bar{\cal Z} \rangle$. 

In the former case, we get 
\begin{equation}
\begin{aligned}
&\langle {\cal Z}^j(x_1)\, \bar{\cal Z}^j(x_2)\,\varphi^a(x_3)\, \varphi^b(x_4)\rangle
=\delta^{ab} \int D\sigma\, j! (2 G_{12})^j G_{34}\, e^{-\frac{N}{2} \log\det\left(-\partial^2+\sigma\right)}\\
&\approx \delta^{ab}\,\frac{j!\, (2 {\cal N})^j C_{\phi}}{|x_{12}|^{2\Delta_j} x_{34}^{d-2}}\,F(X,Y)\,.
\end{aligned}
\end{equation}
Here we have used (\ref{G-final}), and we have defined the function of conformal cross ratios
\begin{equation}
\begin{aligned}
F(X,Y) = \left(\frac{d}{2}-1\right) \int_0^{\infty} dt \,\frac{(Y e^t - X e^{-t}) J_0 \left( \sqrt{c_{\sigma} t(t + \ell)} \right)} {\left( 1 + X (e^{-t}-1) + Y(e^t-1)\right)^{\frac{d}{2}}}\,,\qquad \ell= \log\left(\frac{Y}{X}\right)\,,
\end{aligned}
\end{equation}
where as usual $c_{\sigma}$ is fixed in terms of $\hat{j}$ by (\ref{saddle-csol-extra}), and the conformal cross ratios are
\begin{equation}
X = \frac{x_{13}^2 x_{24}^2}{x_{12}^2 x_{34}^2}\,,\qquad 
Y = \frac{x_{14}^2 x_{23}^2 }{x_{12}^2 x_{34}^2}\,.
\label{XY-cross}
\end{equation}

In a similar way, we find
\begin{equation}
\begin{aligned}
&\langle {\cal Z}^j(x_1)\, \bar{\cal Z}^j(x_2)\, {\cal Z}(x_3)\, \bar{\cal Z}(x_4) \rangle \\
&=j! 2^{j+1}\, \int D\sigma\, \left[G_{12}^j G_{34}+j G_{12}^{j-1}G_{14}G_{23}\right] 
e^{-\frac{N}{2} \log\det\left(-\partial^2+\sigma\right)}\,
\end{aligned}
\end{equation}
At large $N$ with $j/N$ fixed, we again simply evaluate all propagators at the saddle point $\sigma_*$. Note that in the limit we consider, the first term in the square bracket is subleading compared to the second term, due to extra factor of $j=N\hat{j}$ in front of the latter. From (\ref{Gx1y})--(\ref{Gx1x2}), we have
\begin{equation}
\frac{G_{14} G_{23}}{G_{12}} = \frac{{\cal N}}{x_{34}^{d-2}} 
X^{\frac{\sqrt{(d/2-1)^2+c_{\sigma}}-(d/2-1)}{2}}\, Y^{-\frac{\sqrt{(d/2-1)^2+c_{\sigma}}}{2}}\,.
\end{equation}
So the four-point function, to leading order at large $N$ with $j/N$ fixed, is
\begin{equation}
\begin{aligned}
&\langle {\cal Z}^j(x_1)\, \bar{\cal Z}^j(x_2)\, {\cal Z}(x_3)\, \bar{\cal Z}(x_4) \rangle \\
&\approx \frac{2^{j+1}\,j!\, {\cal N}^j\, C_{\phi}}{x_{12}^{2\Delta_j} x_{34}^{d-2}}\, 
\frac{N\hat j}{\sqrt{(d/2-1)^2+c_{\sigma}}} X^{\frac{\sqrt{(d/2-1)^2+c_{\sigma}}-(d/2-1)}{2}}\, Y^{-\frac{\sqrt{(d/2-1)^2+c_{\sigma}}}{2}}\,.
\label{4pt-Z}
\end{aligned}
\end{equation}

Let us make a consistency check of this result with the OPE expansion, in the channel $13\rightarrow 24$. To extract the OPE data in this limit, it is convenient to recast (\ref{4pt-Z}) as (see e.g. \cite{Dolan:2000ut})
\begin{equation}
\begin{aligned}
\langle {\cal Z}^j(x_1)\, \bar{\cal Z}^j(x_2)\, {\cal Z}(x_3)\, \bar{\cal Z}(x_4) \rangle = 
\frac{2^{j+1}\,j!\, {\cal N}^j\, C_{\phi}}{(x_{13} x_{24})^{\Delta_j+\Delta_{\phi}}}
\left(\frac{x_{34}}{x_{12}}\right)^{\Delta_j-\Delta_{\phi}}\,{\cal G}(X,Y)\,,
\end{aligned}
\end{equation}
where, comparing with (\ref{4pt-Z}), remembering $\Delta_{\phi}=d/2-1+O(1/N)$, and using the definition of the cross ratios in (\ref{XY-cross}), we have
\begin{equation}
{\cal G}(X,Y) = \frac{N\hat j}{\sqrt{(d/2-1)^2+c_{\sigma}}} X^{\frac{\Delta_j}{2}+\frac{\sqrt{(d/2-1)^2+c_{\sigma}}}{2}}\, Y^{-\frac{\sqrt{(d/2-1)^2+c_{\sigma}}}{2}}\,.
\label{GXY}
\end{equation}
This function should have the OPE expansion ${\cal G}(X,Y)=\sum_{\Delta,s} a_{\Delta,s}^2 \, X^{(\Delta-s)/2}g_{\Delta,s}(X,Y)$, where the sum is over operators of dimension $\Delta$ and spin $s$ that appear in the $13\rightarrow 24$ channel, $a^2_{\Delta,s}$ are squared OPE coefficients, and $g_{\Delta,s}(X,Y)$ are the conformal blocks (normalized such that $g_{\Delta,s}(X,Y)=1+\ldots$ for $X\rightarrow 0, Y\rightarrow 1$). 

In the limit $X\rightarrow 0$, $Y\rightarrow 1$, the leading contribution should come from a scalar operator of charge $j+1$ that appears in the OPE of ${\cal Z}^j$ and ${\cal Z}$. Comparing (\ref{GXY}) with the OPE expansion in this limit, we see that the dimension of the exchanged operator of charge $j+1$ should satisfy
\begin{equation}
\Delta_{j+1} -\Delta_j = \sqrt{(d/2-1)^2+c_{\sigma}}\,.
\label{Del-OPE}
\end{equation}
This is precisely as expected. Indeed, writing $\Delta_j = N h(\hat{j})$, we have in the large $N$ limit
\begin{equation}
\Delta_{j+1}-\Delta_j = N h\left(\hat{j}+\frac{1}{N}\right)-N h(\hat{j}) = \frac{\partial h}{\partial \hat{j}}\,.
\end{equation}
On the other hand, from (\ref{Delta-final}) and (\ref{saddle-csol-extra}), we see that 
\begin{equation}
\frac{\partial h}{\partial \hat{j}} = \sqrt{(d/2-1)^2+c_{\sigma}}\,,
\end{equation}
in agreement with (\ref{Del-OPE}). From (\ref{GXY}) we can also read off the squared OPE coefficient
\begin{equation}
a^2_{j,1,j+1} = \frac{N\hat j}{\sqrt{(d/2-1)^2+c_{\sigma}}}\,.
\end{equation}
This is in precise agreement with the result (\ref{a-final}) derived earlier, setting $j_3=1$, $q=0$. 

\section{Conclusion}
\label{Concl}
In this paper we have studied large charge operators in the large $N$ critical $O(N)$ model in general $d$, in the limit where the charge $j$ goes to infinity with $\hat{j}=j/N$ fixed. In particular, we have obtained the scaling dimensions to leading order at large $N$ and arbitrary $\hat{j}$, as well as the 3-point and 4-point functions involving two large charge operators. In the range $4<d<6$, we have observed an interesting transition from real to complex scaling dimensions at a critical value of the ratio $j/N$, which we view as a manifestation of the instability of the interacting $O(N)$ model in $4<d<6$ that is not captured by the ordinary $1/N$ perturbation theory. 

There are several extensions of our results that would be worth pursuing. For example, a natural further step would be to compute the subleading corrections to the scaling dimensions and other observables in the large charge limit we considered. For instance, the order $N^0$ correction can be computed by including the one-loop determinant arising from the quantum fluctuations around the semiclassical saddle point we found in Section \ref{2pt-saddle}. It would be interesting to evaluate such correction to $\Delta_j$ explicitly for arbitrary $\hat{j}$ and $d$. It would be also useful to extend the calculation of correlation functions to the case of more than two heavy operators. For instance, deriving the 3-point function coefficients in the ``heavy-heavy-heavy" configuration would be an interesting and non-trivial problem. 

It would be also interesting to further investigate the instability of the theory in $4<d<6$, and in particular understand the relation between the complex dimensions of the large charge operators that we found here and the imaginary part of the thermal free energy computed in \cite{Petkou:2018ynm, Giombi:2019upv}.  

Another natural direction would be to see if the methods we used in this paper can be extended to other kinds of operators with large quantum numbers. For example, one could consider other large representations of $O(N)$, or operators with spin $s$ in the large $N$ limit with $s/N$ fixed. In the case of the critical $O(N)$ model, or its generalizations involving Chern-Simons gauge theory, this may have interesting applications to the duality \cite{Klebanov:2002ja, Giombi:2012ms, Giombi:2016ejx} with Vasiliev higher spin theory in AdS \cite{Vasiliev:1990en, Vasiliev:2003ev}. Since the bulk coupling constant is identified with $1/N$, the CFT states with quantum numbers of order $N$ should be related to non-trivial classical solutions of the bulk higher-spin theory. 

\section*{Acknowledgments}

We are grateful to Igor Klebanov for useful discussions and comments. This work was supported in part by the US NSF under Grant No.~PHY-1914860. Some of the results presented here are from Jonah Hyman's Princeton University Senior Thesis (May 2020).

\bibliographystyle{ssg}
\bibliography{mybib}

\end{document}